\documentclass{article}

\usepackage[utf8]{inputenc}

\usepackage{amsfonts}
\usepackage{amsthm}
\usepackage{amsmath}
\usepackage{graphicx}

\usepackage [english]{babel}
\usepackage [autostyle, english = american]{csquotes}
\MakeOuterQuote{"}

\usepackage {natbib}
\shortcites{campis, ada, jansson, Anderwald}

\usepackage{algorithm2e}
\usepackage{hyperref}

\newtheorem{lemma}{LEMMA}

\usepackage[usenames]{color}         

\title{Bayesian Experimental Design for Oral Glucose Tolerance Tests (OGTT)}
\date{}
\author{Nicolás  E. Kuschinski$^a$, J. Andrés Christen$^a$, Adriana Monroy$^b$, Silvestre Alavez$^c$\\
$^{a}${\em CIMAT, Guanajuato, Gto., Mexico}\\
$^{b}${\em Hospital General, Mexico City, Mexico}\\
$^{c}${\em Universidad Autónoma Metropolitana, Unidad Lerma, Mexico City, Mexico}}

\begin{document}

\maketitle

\begin{abstract}
OGTT is a common test, frequently used to diagnose insulin resistance or diabetes, in which a patient's blood sugar is measured at various times over the course of a few hours. Recent developments in the study of OGTT results have framed it as an inverse problem which has been the subject of Bayesian inference. This is a powerful new tool for analyzing the results of an OGTT test, and the question arises as to whether the test itself can be improved. It is of particular interest to discover whether the times at which a patient's glucose is measured can be changed to improve the effectiveness of the test. The purpose of this paper is to explore the possibility of finding a better experimental design, that is, a set of times to perform the test. We review the theory of Bayesian experimental design and propose an estimator for the expected utility of a design. We then study the properties of this estimator and propose a new method for quantifying the uncertainty in comparisons between designs. We implement this method to find a new design and the proposed design is compared favorably to the usual testing scheme.
\end{abstract}

\section{Introduction}

Diabetes is a serious and potentially fatal illness that is on the rise and is expected to affect over 4\% of the worldwide population by the year 2030 \citep{ada,ama}. Diabetes occurs when the pancreas cannot produce enough insulin (a hormone which lowers blood glucose), or when the body is unable to efficiently use the insulin it produces, thereby reducing the effectiveness of the body's blood sugar regulation.

While type 1 diabetes is usually diagnosed very early in life, and is related to genetic disorders, the same cannot be said for type 2 diabetes, which is a far more common ailment that is acquired at an older age. Unlike type 1 diabetes, which comes on suddenly and produces obvious symptoms, type 2 diabetes usually develops without any noticeable symptoms initially. Therefore it can often go undiagnosed for years. With a timely diagnosis and proper treatment, type 2 diabetes can change from a serious health risk to a relatively mild condition. Early diagnosis can also serve to identify patients who are at risk of developing type 2 diabetes and take steps to prevent this from occurring \citep{ama}.

In order to administer treatment or take preventative measures, it is of vital importance to have a good means of diagnosis. One common technique for diabetes diagnosis is the Oral Glucose Tolerance test (OGTT). To perform this test, a patient arrives after a night of fasting and has his/her blood glucose measured (in \textit{mg} of glucose per \textit{dl} of blood). The patient is then asked to drink a 75g glucose concentrate. Blood glucose is measured again at various times (typically over the course of two hours) and these measurements are used to infer the body's ability to regulate sugar \citep{ama, jansson, mayer}.

We aim to improve the ability of OGTT tests to provide a more complete accurate analysis, investigating alternative experimental designs. The primary design variable of interest is the times $t_i$ at which measurements of blood glucose are taken. In order to find a particularly useful set of times, we use the theory of Bayesian experimental design to analyze a recently proposed model for the analysis of OGTT tests.

This paper is organized as follows: Section~\ref{model} presents a new dynamic model for analyzing OGTT tests and section~\ref{infer} shows how this model is used to perform Bayesian inference on OGTT data. Section~\ref{improve} presents the main question of this paper: How to use this mathematical model to redesign and improve the OGTT test itself. Section~\ref{design} presents the theory of Bayesian experimental design while in section~\ref{alg} we offer a new algorithm to decide between designs and section~\ref{prior} discusses the issue of prior selection for diagnosis of diabetes using OGTT tests, particularly in the context of experimental design. Finally, section~\ref{resul} presents the results of using the algorithm to study the problem of design for OGTT tests.


\section {Modeling blood glucose}\label{model}

Although the idea of using mathematical models to analyze OGTT results is not new \citep{jansson}, the current medical practice to interpret the results of an OGTT test is based on extremely simple guidelines. Blood sugar measurements are inspected by hand, and patients are usually diagnosed as diabetic if measurements are elevated above a certain threshold (usually thought of as $\geq200 mg/dl$). Commonly used markers include the maximum measurement, the last measurement, and the average \citep{mayer}. These markers are all extremely basic and, in fact, all of them completely ignore the temporal element of the data. A proper analysis should take into account the fact that the data correspond to repeated measurements of a process over time.  In order to properly use temporal information,
we use an ODE system to model the process.  This represents a dynamic minimal model for OGTT \citep[see also][]{campis}.
This model is presented in the following section.

\subsection {The dynamic model}\label{sec.dynmodel}

The dynamic model which has been proposed to analyze OGTT results is the following

\begin{align}
\frac{dG}{dt} &= L-I+\frac{D}{\theta_2}\label{eq:dyn1}\\
\frac{dI}{dt} &= \theta_0(G-G_b)^+-\frac{I}{a}\label{eq:dyn2}\\
\frac{dL}{dt} &= \theta_1(G_b-G)^+-\frac{L}{b}\label{eq:dyn3}\\
\frac{dD}{dt} &= -\frac{D}{\theta_2}+\frac{2V}{c}\label{eq:dyn4}\\
\frac{dV}{dt} &= -\frac{2V}{c}\label{eq:dyn5}.
\end{align}

Here $G$ represents blood glucose in $mg/dl$, $I$ is insulin, $L$ is glucagon (a hormone produced by the pancreas that interacts with the liver to raise blood glucose), $D$ is the glucose in the digestive system and $V$ is the glucose which has not yet entered the digestive system (it has not yet been ingested). $a,b,c$, and $G_b$ are assumed to be known constants. The processing of each patient's blood glucose is dependent on four parameters: $\theta_0$, $\theta_1$, $\theta_2$ and $G(0)$ which represent insulin sensitivity, glucagon sensitivity, glucose digestive system mean life, and glucose level at the start of the test, respectively. The values of $a,b,c$, and $G_b$ are taken from the literature, \cite{Anderwald}. 

This model follows a simple three compartment and double feedback logic, as we explain next. As the patient consumes the glucose concentrate, glucose moves into the digestive system (see (\ref{eq:dyn5}) and (\ref{eq:dyn4})). Once in the digestive system, glucose enters the blood at a rate determined by $\theta_2$ (see (\ref{eq:dyn4}) and (\ref{eq:dyn1})). As blood glucose rises, insulin $I$ is produced to regulate this with a level of efficiency that depends on $\theta_0$ (see (\ref{eq:dyn2}) and (\ref{eq:dyn1})). When glucose goes below the base level $G_b$ then glucagon $L$ is produced to regulate it with a level of efficiency dependent on $\theta_1$ (see (\ref{eq:dyn3}) and (\ref{eq:dyn1})). 

\begin{figure}
\includegraphics[scale=0.35]{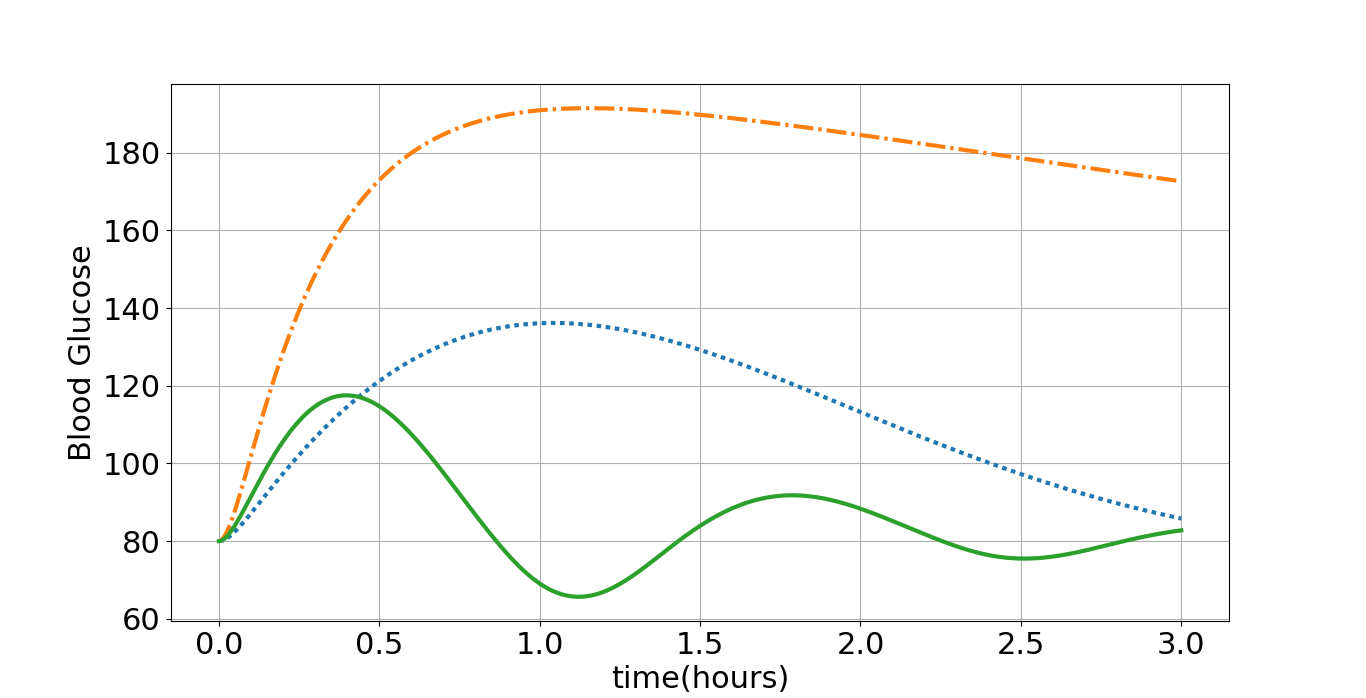}
\caption{Three typical OGTT curves, reproduced by the model. The horizontal axis is time and the vertical axis is blood glucose.
The dotted line is a healthy patient ($\theta_0 = 2.15, \theta_1 = 1.3, \theta_2 = 0.8$),
the broken line is a diabetic patient ($\theta_0 = 0.2, \theta_1 = 3.52, \theta_2 = 0.3$)
and the solid line is a patient who produces an unusually large amount of insulin ($\theta_0 = 15.3, \theta_1 = 31.35, \theta_2 = 0.6$),
leading to glucose oscillation; a situation which has turned out not to be rare.}
\label{fig:dynamic}
\end{figure}

Figure~\ref{fig:dynamic} shows the behavior of the model with various values of $\theta_0$, $\theta_1$ and $\theta_2$. The vertical axis is blood glucose and the horizontal axis is time. The three lines represent three typical scenarios for how a patient's body might react to glucose. The dotted line is a healthy patient. This patient arrives with his/her blood sugar in equilibrium; his/her glucose rises when he/she ingests the glucose solution and then his/her body regulates his/her blood glucose which returns to normal after about 3 hours. The broken line is a diabetic patient, whose body does not adequately regulate blood glucose, and even after 3 hours his/her blood glucose still has not returned to initial levels. The solid line represents a patient producing an unusually large amount of insulin resulting in that the patient's blood sugar then goes below his/her base condition and then must be raised with glucagon production, leading to an oscillation. Previous OGTT studies would have classified this patient as "normal". Upon fitting the model to numerous patients we have found that this kind of situation is not at all rare.

All of these curves eventually behave as damped harmonic oscillators, but this oscillation is usually not visible in the first few hours unless $\theta_0$ and $\theta_1$ are relatively high, as it is the case in the third scenario just described.

\section {Inferring OGTT curves from data}\label{infer}

The data collected in an OGTT test are measurements of $G(t)$ at various times. For instance, for an actual patient,
at times $t=$0:00, 0:30, 1:00, 1:30 and 2:00 hours we obtained glucose measurements of $y= 81,156,141,102, 89~ mg/dl$ (these data, with measurements taken every half hour).  The intent is to use these data to infer the parameters $\theta_0, \theta_1$ and $\theta_2$.
This is an "inverse" or "UQ" problem, that is, an inference problem in which the regressor is
driven by a system of differential equations, see \cite{Fox2013}.

In order to perform inference, we assign priors to the parameters (see section~\ref{prior} for more details) and assume the data $y$
follows a independent additive Gaussian error model, namely
$$
y_i=G(t_i)+\epsilon_i
$$
where $\epsilon_i\sim\mathcal{N}(0,\sigma)$, where $i=1,2, \ldots , d$. We assume $\sigma$ is fixed at $\sigma=5$ as recommended by the laboratory staff. 
Thus, we write the likelihood as
$$
f( y | \theta ) \propto \prod_{i=1}^d e^{-\frac{ y_i - G( t_i | \theta)}{2 \sigma^2}} .
$$
Using Bayes's theorem, we find that our posterior distribution is
$$
\pi(\theta | y)\propto f(y | \theta)\pi(\theta) .
$$

\begin{figure}
\includegraphics[scale=0.3]{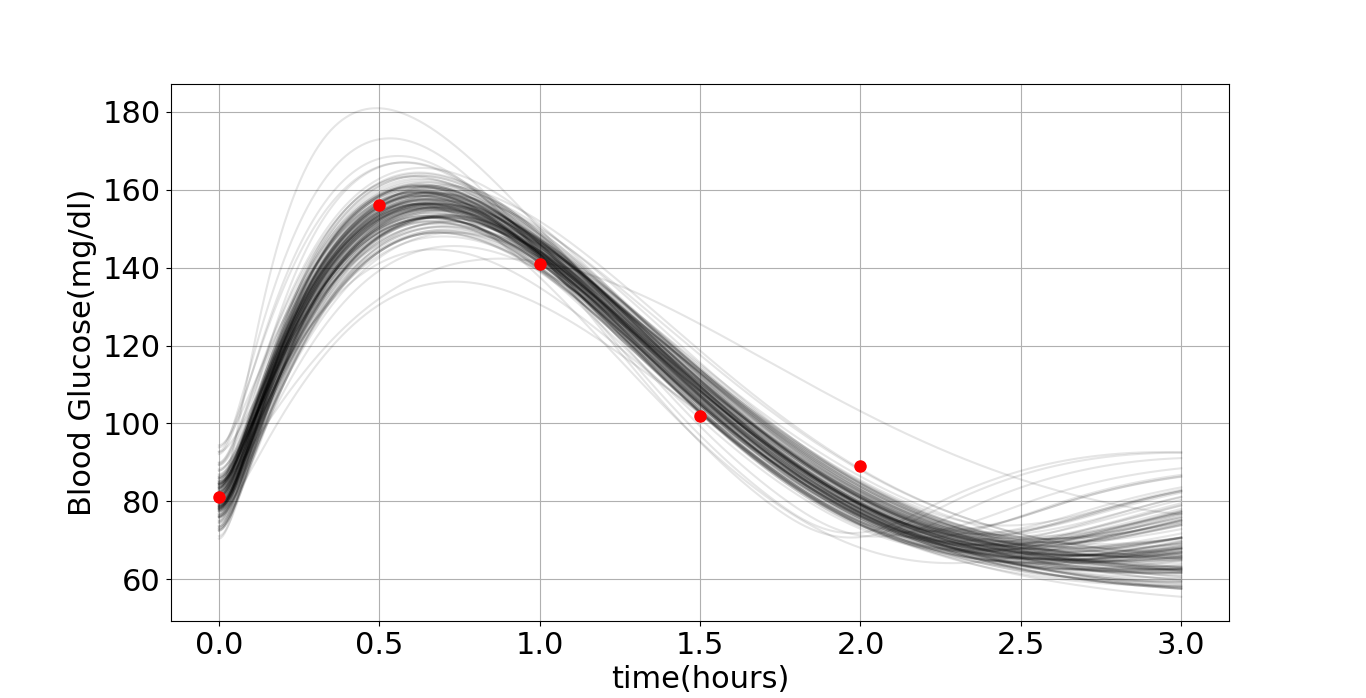}
\caption{The result of OGTT inference on a healthy patient. The dots are the data points and each grey curve is the resulting curve from one element of a posterior sample.}
\label{fig:inference}
\end{figure}

To perform inference on this model, an MCMC simulation is used to obtain a posterior sample.  In order to calculate $G(t)$, the forward map is solved numerically using the LSODA package for ordinary differential equations and to perform the MCMC the t-walk package \citep{christen} is used. Figure~\ref{fig:inference} shows a posterior sample from a real (healthy) patient, presented above.
The dots are the data points and the grey curves are the resulting OGTT curves from each
element of the posterior sample which is obtained. As we can see, at least for this patient, the model appears to fit the data quite well.

\section {Improving OGTT tests} \label{improve}

The use of a dynamic model to analyze the results of OGTT tests represents a significant potential improvement over
the current guidelines since it attempts to describe how the patient's body handled the ingested glucose over the duration of the test.  The use of this ODE model may help to improve the OGTT data analysis, once these data have been obtained.
An additional point of interest
is to design the times $t_i$ at which glucose is measured, to make the most of the information gained by the test.
Common practice is to measure glucose at $t_1=0$ (arrival), $t_2=1$ and $t_3=2$ hours.  Note that, in section~\ref{resul} we use data taken at more frequent intervals (up to every 15 min).  These more frequent measurements are uncommon, but are occasionally taken for research purposes (OGTTs performed by AM).

It is not difficult to come up with \textit{bad} sets of times at which to measure. For instance, if we look back to figure~\ref{fig:dynamic} we can see that the red and blue lines cross around time $t=0:30$ hours and that they converge again around time $t=3:00$ hours. If we only measure at times $t_1=0:00$, $t_2=0:30$ and $t_3=3:00$ then it will not be possible to distinguish between these two significantly different scenarios. It is also worth noting that, without referring back to actual curves, this set of times is \textit{not} obviously problematic.

It is clear that any testing scheme can be improved by increasing the number of times at which a measurement is taken (the patient from figure~\ref{fig:inference} had samples taken at $t=$ 0:00,  0:30,  1:00, 1:30, 2:00 hours, but it would be desirable to know how many times are enough as well as what times these should be.

We will use Bayesian experimental design to propose a new set of times with the aim of improving the test.

\section {Experimental design} \label{design}

Most of statistics is concerned with inference from collected data, ignoring the issue of how the data is collected to begin with. This is, overall, a significant omission, since the quality of inference typically depends heavily on the quality of the data. Experimental design is the use of statistical techniques to improve the quality of the data that will be collected, and thus improve the quality of the resulting inference \citep{berger}.

Unlike most areas of statistics, experimental design is concerned with what happens \textit{before} any data becomes available. In Bayesian statistics in particular, the information available before the presence of data is encoded in the prior distribution. While Bayesian inference is concerned with the results of studying the interaction of the prior distribution and data, experimental design is primarily interested in studying the properties of the prior distribution itself, considering what it says regarding the data that might be obtained when it is actually collected.

In the case of OGTT tests, a design is $d=\left(t_1,\ldots t_n\right)$, the times at which blood samples are drawn for testing.

For a given design $d$, it is common practice to write $\pi(y|\theta,d)$ as the likelihood function for the data $y$ and the parameter $\theta$ when using the design $d$, see for example \cite{youssef} for one case where this is used. This is not actually a conditional probability in the strict sense of the word: $d$ is not a random variable, and $\pi(d)$ does not exist. This practice is therefore notational abuse. It is, however, standard, and here we conform to this notation.

\subsection{The main idea: Utility functions}

Let $d_1$ and $d_2$ be designs which we want to compare. We define a utility function $u$, which assigns a value to the result of an experiment. In the most general sense, the utility is a functional from the space of posterior distributions to $\mathbb{R}$, however we need not worry about this representation since the posterior distribution is determined by finite dimensional data. It is therefore possible to write the utility as a function of data directly $u(y)$. Note that since $u(y)$ depends on random data, a priori it is itself a random variable.

A utility function may be, for instance, equal to minus the posterior variance of one component of the parameter, or minus some norm of the difference between the predictive distribution and the true distribution which generates data (assuming such a thing exists), etc. Another choice is to use the K-L divergence between the prior and posterior distributions, the idea being that the bigger the difference between these distributions, the more information was acquired from the data; see \cite{youssef,meeker,chrisbuck,kathry,farminde,steve,alexanderian,weaver,solonen} for several examples of utility functions and Bayesian desing problems

We write $u(y|d)$ as the utility for the data $y$ collected using an experimental design $d$. Our selection between $d_1$ and $d_2$ is based on which of these designs maximizes the expected utility given our prior distribution, namely
$$
U(d|\pi)=\int u(y|d)\pi(y|d)dy .
$$
The distribution under which the expectation is calculated is the predictive prior distribution of the data under the design $d$:
$$
\pi(y|d)=\int\pi(y|\theta,d)\pi(\theta)d\theta.
$$
While it is common to simply write $U(d)$, note that $U(d)$ is also dependent on the prior $\pi(\theta)$.


The main goal of experimental design is to find good designs, which means we want a design $d$ such that $U(d)$ is as high as possible. Unfortunately, in our case, $U(d)$ is not tractable. The difficulty lies not only in calculating the integral $\int u(y|d)\pi(y|d)dy$ since even with specified data $y$, it is usually not possible to calculate $u(y)$ exactly.

The real difficulty in this experimental design, and a common issue in Bayesian experimental design in general, lies in optimizing a function which cannot be evaluated exactly. The approach we take here is to find good Monte Carlo estimators for $U(d)$ and use them for design comparisons on a comprehensive discrete grid of possible designs.  The optimization is then taken, not over the continuous time space, but only over a discrete space of 15 min intervals, proceeding by semi-brute force maximization.  We explain the details of this approach in section~\ref{alg} and onwards. 
Meanwhile, in the next section we explain briefly other common approaches for implementing Bayesian experimental designs and
why these are not suitable for our design problem.

\subsection{Other approaches}

The number of published papers on Bayesian experimental design is very small relative to the amount of research done in Bayesian statistics and also very small relative to the amount of research done in experimental design in general. Nonetheless, some techniques have been proposed to optimize design parameters. All of these approaches are based on the idea of finding
$$
d^*=argmax_d U(d)
$$
using some kind of optimization algorithm.

Most traditional optimization techniques are not useful with a function as poorly understood as $U(d)$ \citep[classical optimization techniques have been attempted using random estimations of $U(d)$,][ but this does not have adequate theoretical justification]{farminde}, so specialized approaches must be taken. There are two main ideas to try to circumvent this problem.

\begin {enumerate}
\item{\textbf{Asymptotic estimations of $U(d)$:}} A well known result of Bayesian statistics states that under certain regularity conditions, the posterior distribution approaches a Gaussian as sample size goes to infinity. If we assume that the sample size is large, then for certain utility functions it is possible to calculate the asymptotic value of $U(d)$. Then $U(d)$ is optimized in the asymptotic regime \citep{meeker,kathry,steve}.

While this approach may be reasonable under some circumstances, it is worth noting that the entire problem of experimental design is most interesting precisely when sample sizes are \textit{small}. Under most circumstances, increasing the sample size is an easy way to improve a design, and the need for a well-designed experiment arises only when circumstances indicate that large sample sizes are impossible to begin with.

\item{\textbf{Stochastic approximation:}} There is a class of numerical techniques called \textit{stochastic approximation} techniques which deal with functions that cannot be directly measured. Some of these techniques are used for optimization, and these have been used to attempt to optimize $U(d)$ without requiring a large sample size. By far the most commonly used of these techniques is the Robbins-Monro algorithm \citep{robbins,youssef,duflo}.

Most stochastic approximation techniques (including Robbins-Monro) require an unbiased estimator of the gradient $\frac{\delta U}{\delta d}$. With certain utility functions it is possible to obtain this estimator, but this is not universal.

Although there are some derivative free stochastic approximation techniques \citep{duflo}, a more serious issue than the requirement of gradients is the fact that all of these algorithms only perform local optimization. In fact, to our knowledge, the convergence of stochastic optimizers has only been proven for strongly convex functions \citep{duflo}. If $U(d)$ is a well-behaved strongly convex function then this is not an issue, but in our case we have no reason to believe that our expected utility $U(d)$ belongs to such class.

Furthermore, stochastic approximation algorithms are based on simulations and estimations. They are therefore subject to error based simply on the randomness of the estimators. While avoiding this kind of error altogether is impossible, it would be extremely desirable to control - or at least quantify - the uncertainty in our eventual conclusions as a result of these errors.
\end {enumerate}

Recently an alternative has been proposed which approximates the utility function using Gaussian processes. This estimation is not asymptotic with sample size and may be a more robust alternative to asymptotic estimations \citep{alexanderian, weaver}.

We propose a different alternative which does not have these problems, but which suffers from a different set of limitations. Rather than attempting to find an optimum design, we simply propose a good way to decide between any two designs, and then perform many comparisons, optimizing by semi brute force. This is generally not a good technique for optimization, but in this case its use is warranted since it allow us to perform comparisons that do not depend on sample size or any special properties of the function $U$, to avoid bad local maxima, and also to control and quantify the uncertainty in our conclusions. While we may not be able to reach any definite conclusion about what design is actually the best, we will be able to achieve arbitrarily high confidence in our claims regarding what designs are good.

\subsection{An unusual generalization}\label{sec:multiple_priors}

We make a generalization to the usual scheme of experimental design, which is to allow for two different priors over the same parameter: One for design ($\pi_{\mathcal{D}}$) and one for inference ($\pi_{\mathcal{I}}$). This generalization is admittedly unusual: It is not clear that anyone would ever \textit{want} to allow $\pi_{\mathcal{D}}$ and $\pi_{\mathcal{I}}$ to be different. For now we limit ourselves to indicate that this generalized problem is indeed well-defined. The motivation for this generalization is discussed in section~\ref{prior}, where we see how, in some situations, this may indeed be desirable.

Accordingly, let $\pi_{\mathcal{I}}(\theta)$ be the prior which is used for performing inference, and let $\pi_{\mathcal{D}}(\theta)$ be the prior used to design the experiment. If $\pi_\mathcal{D}=\pi_\mathcal{I}$ then we have the usual problem, as described above. The general function which we wish to optimize is $U(d|\pi_\mathcal{D})$ which can be written as
$$
U(d|\pi_\mathcal{D})=\int u_{\mathcal{I}}(y|d)\pi_{\mathcal{D}}(y|d)dy ,
$$
where the utility $u_{\mathcal{I}}(y|d)$ is calculated using the posterior
distribution generated by the inference prior $\pi_{\mathcal{I}}$.

For the remainder of this paper we will write $U(d|\pi_{\mathcal{D}})$ as simply $U(d)$ and $u_{\mathcal{I}}(y|d)$ as simply $u(y|d)$. A design will be selected in a way that works regardless of whether or not the priors are different.

\section{The algorithm for design selection}\label{alg}

\subsection{Estimation of $U(d)$}

We require some technique for estimating (but not necessarily directly calculating) $u(y|d)$ for any given $y$ and $d$. Note that unless $y$ is fixed, $u(y|d)$ depends on random data $y$, and is hence a random variable itself. Since $u$ is typically a functional of the posterior distribution, we assume that we can estimate $u(y|d)$ with a posterior sample $\vartheta$ produced by some posterior sampling algorithm such as an MCMC chain. We call this estimator $\hat{u}(y|d)(\vartheta)$. Note that $\hat{u}$ depends on the random data $y$ and also on the random posterior sample $\vartheta$.
We have two requirements on $\hat{u}$. The first is that it is unbiased for fixed $d$ and $y$ (this requirement can be relaxed to allow for asymptotically unbiased estimators, but this comes at a cost, see section~\ref{propt1yt2}), and second that it have finite second moment. In other words, that $\mathbb{E}_{(Y|d)_\mathcal{D}}[\hat{u}(y|d)^2]<\infty$. In our case, $u(y|d)$ is minus the mean squared error of $G(t)$ integrated from $t=0$ to $t=3$ hours. This can be estimated using Monte Carlo samples, as described in section~\ref{resul}

We now propose the following sampling algorithm for estimating $U(d)$ for a given $d$:

\vspace {1cm}
\begin{algorithm}[H]
Fix two constants $T_1$ and $T_2$:

\For {$i$ from 1 to $T_1$}{
Sample data $y^{(i)}$ from the predictive prior $\pi_{\mathcal{D}}(Y|d)$\;
Generate a sample $\vartheta^{(i)} = \{\theta^{(i,j)}:j=1\ldots,T_2\}$ from the posterior $\pi_\mathcal{I}(\theta|y^{(i)},d)$\;
Calculate $\hat{u}_i=\hat{u}(y^{(i)}|d)(\vartheta^{(i)})$\;
}
Calculate $\hat{U}(d)=\frac{1}{T_1}\sum_i\hat{u}_i$
\end{algorithm}

\vspace{1cm}
We now calculate the expected value of our estimator $\mathbb{E}_\mathcal{D}(\hat{U}(d))$:
\begin{align*}
\mathbb{E}_\mathcal{D}(\hat{U}(d)) &= \frac{1}{T_1}\sum_i\mathbb{E}_{(Y|d)_\mathcal{D}}(\hat{u}_i)\\
    &= \frac{1}{T_1}\sum_i\mathbb{E}_{(Y|d)_\mathcal{D}}\hat{u}(y_i|d)(\vartheta^{(i)}))\\
&= \mathbb{E}_{(Y|d)_\mathcal{D}}\hat{u}(y|d)(\vartheta)\\
\end{align*}
where $\vartheta$ is a random variable with the same distribution as any $\vartheta_i$

    Now we observe that since $\hat{u}(y^{(i)}|d)(\vartheta)$ is unbiased then
$$
\mathbb{E_{\mathcal{D}}}(\hat{U}(d))=U(d)
$$
so $\hat{U}(d)$ is an unbiased estimator.

Moreover, observe that $\hat{U}(d)$ is an average of \textit{iid} random variables, each of which is distributed as $\hat{u}(y|d)$, which has finite second moment. Hence, $\hat{U}(d)$ is subject to the central limit theorem, so as $T_1\rightarrow\infty$ we have
$$
\mathbb{P}\left(\frac{\hat{U}(d)-U(d)}{\sqrt{var(\hat{U}(d))}}<\alpha\right)\rightarrow\Phi(\alpha) .
$$
Now $var(\hat{U}(d))=var(\frac{1}{T_1}\sum_i\hat{u}_i)=\frac{1}{T_1}var(\hat{u}_i)$. We can estimate $var(\hat{u}_i)$ with its sample variance, and arrive at a normal asymptotic distribution for $\hat{U}(d)$.

Similar estimators have been proposed in the past \citep{chrisbuck, farminde} but the properties of the estimators (such as their asymptotic distribution) were not studied. In the following section we explain how this distribution can be used to quantify uncertainties in comparisons between designs.

\subsection{Numerically deciding between $d_1$ and $d_2$} \label{decide}

Now that we are able to estimate $U(d_1)$ and $U(d_2)$ the simplest idea is to select the design with the higher estimator. This is not satisfactory, however, unless we have a proper way of controlling the uncertainty in this choice. Since our estimators depend on random simulations, we want to be certain that the difference between these estimators corresponds to an actual difference in the expected utility of the experiments and is not solely the result of the random nature of the estimators.

We have an asymptotic distribution for $\hat{U(d_1)}$ and $\hat{U(d_2)}$. Hence, so long as $T_1$ is sufficiently large, we can consider the comparison of two expected utilities to be a comparison of the means of two normally distributed random variables with known variance. (Technically the variance is unknown, but if $T_1$ is large enough this is not a problem. Theoretically, errors in variance estimation can be handled using a $t$ statistic rather than a normal statistic, but the distribution of the statistic depends on the sample size. Furthermore, for large samples, the resulting $t$ distribution is almost identical to a normal one anyway.) This is a well-studied classical problem.

Assume, with no loss of generality, that $U(d_1)\leq U(d_2)$. Now we fix a value $0<\alpha<1$ and we wish to make sure that the probability of wrongly concluding that $U(d_2)>U(d_1)$ is at most $\alpha$. This can be done by considering the variable
$$
Z=\frac{\hat{U}(d_1)-\hat{U}(d_2)}{\sqrt{var(\hat{U}(d_1))+var(\hat{U}(d_2))}} .
$$
$Z$ is asymptotically normally distributed with mean 0 an variance 1 \citep{degroot}, so we can conclude that $U(d_1)<U(d_2)$ if $Z$ is less than the $\alpha/2$ quantile of a standard normal distribution.

It is still possible that this problem will not be completely solved since testing for $U(d_1)<U(d_2)$ and also testing for $U(d_2)<U(d_1)$ may both produce inconclusive results. This does not necessarily mean that the two designs are of equal (or even of approximately equal) expected utility, but rather that the variance of our estimators is still too large to be able to choose with the required degree of certainty. In section~\ref{resul} we discuss how this does not represent a problem in our case, although in some other situations it might become an issue.

If reaching decisive conclusions is required, then it is possible to increase the sample size and test again. This presents a problem; the probability of error when testing repeatedly is greater than the probability of error when testing once since the error could have been committed at any of the tests.  However, it is possible to implement a sequential testing scheme in the style of the Sequential Probability Ratio Test \citep{wald}. The classical form of the Sequential Probability Ratio Test requires knowledge of the power of the test, which is unavailable in our situation, but it can be modified slightly to work in this situation as well.  We have explored some implementations of this idea, but it is not yet clear how to accomplish this task efficiently.

\subsection{How the choice of $T_1$ and $T_2$ affects estimation} \label{propt1yt2}

Note that when we perform sequential testing, the way we reduce the variance of our estimator is to increase $T_1$, but it is also possible to reduce the variance by increasing $T_2$. However, $T_1$ and $T_2$ have very different effects on the distribution of $\hat{U}(d)$.

Our first observation is that increasing $T_2$ only reduces the variance of $\hat{u}(y^{(i)}|d)(\vartheta^{(i)})$ by increasing the size of the sample $\vartheta^{(i)}$, but even if we were able to calculate $u(y^{(i)}|d)$ exactly for each $i$, that still will not reduce $var(\hat{U}(d))$ to zero, since $y^{(i)}$ is still random. In other words, $T_1$ absolutely must be increased to assure that one of the models is eventually selected. Increasing $T_2$, however is not strictly required.
\begin{lemma}
$var(\hat{U}(d))\rightarrow 0$ is assured so long as $T_1\rightarrow\infty$
\begin{proof}
$var(\hat{U}(d))=var\left(\frac{1}{T_1}\sum_i{\hat{u}_i}\right)=\frac{var(\hat{u}_1)}{T_1}\rightarrow 0$
\end{proof}
\end{lemma}

That is, remembering what $T_1$ and $T_2$ are, increasing the number $T_2$ of (MCMC) samples for each posterior
given a simulated sample does not assure that our estimator of the design utility $\hat{U}(d)$ tends to zero.
On the contrary, only the number of simulated samples $T_1$ for the design $d$ needs to increase and $T_2$
could be kept fixed, and possibly low, as we discuss next.

The second observation is that if $T_2$ is unchanged then it is possible to continue the algorithm, drawing more samples from the predictive prior. These can be used to increase $T_1$ without discarding the previous sample. It is not possible to do this if we attempt to increase $T_2$ for the new sample points since altering $T_2$ changes the sample size from which $\hat{u}(y^{(i)}|d)$ is calculated and therefore alters the distribution of the estimator. These simple and useful observations also apply to many similar algorithms but were overlooked by previous authors \citep{chrisbuck,farminde}. 

In general, the effect of $T_1$ and $T_2$ to reduce $var(\hat{U}(d_k))$ depends heavily on the loss function and the model, but the previous two observations make it seem reasonable to suppose that it is a good idea to have $T_2$ be "fast" (of course, it must be a bare minimum large enough to obtain an unbiased estimator $\hat{u}(y|d)(\vartheta)$) and allow $T_1$ to increase dynamically.

Note that if $\hat{u}(y|d)(\vartheta)$ is asymptotically unbiased -- rather than unbiased for finite sample size -- then this does not work equally well. The central limit theorem only states
$$
\mathbb{P}\left(\frac{\hat{U}(d)-\mathbb{E}(\hat{U}(d))}{\sqrt{var(\hat{U}(d))}}<\alpha\right)\rightarrow\Phi(\alpha)
$$
where for unbiased estimators $\mathbb{E}(\hat{U}(d))$ can be replaced by $U(d)$. For asymptotically unbiased estimators, the usefulness of the approximation depends on the quality of the approximation $\mathbb{E}(\hat{U}(d))\approx U(d)$. This in turn depends on $\mathbb{E}_{(Y|d)_\mathcal{D}}(\hat{u}_i)\approx u(y^{(i)})$, and the quality of this approximation depends on the sample which is used to calculate it. That sample is of size $T_2$. Hence, if we intend to use an asymptotically unbiased estimator for the utility, then the quality of our estimation will depend on $T_2$. This is a strong reason to prefer an unbiased estimator if one is available.

\subsection{Special considerations for MCMC type samplers}

The aforementioned method for hypothesis testing does not depend on the technique used to obtain a posterior sample, but in practice the most common method is the use of MCMC algorithms such as the Gibbs Sampler or the Metropolis-Hastings algorithm.

There are two issues which are of particular interest when using MCMC for sampling. The first is the issue of obtaining a proper estimator $\hat{u}(y|d)$. It is worth noting that proximal iterations of an MCMC chain are usually strongly correlated. There has been much debate as to whether an MCMC sample should be "thinned" by taking only one iteration every so often (to avoid correlation of proximal iterations) in the chain for posterior inference or if it is OK to treat the full chain as the posterior sample of interest.

The answer to the thinning question in general depends on what information is desired from the posterior. In this particular case, what is needed is an unbiased estimator for $u(y|d)$. Common cases of unbiased estimators require an \textit{iid} sample, and hence, most of the time the MCMC chain \emph{must} be thinned.

The second issue of interest in an MCMC algorithm relates to burn-in times. When running an MCMC algorithm there are two parameters of note which affect the running time for the posterior estimation: These are the autocorrelation time and the burn-in time. When the reason to generate a posterior sample is to perform inference, the time which is most important to reduce is autocorrelation time, since for a size $m$ sample the algorithm must run through the autocorrelation time $m-1$ times, and the burn-in time only once.

For this form of experimental design, however, large burn-in times can also be very problematic since an MCMC chain must be run $T_1$ times to obtain an estimator of $U(d)$ for a single design. If burn-in times are significant then this can be a problem. Luckily in this situation it is possible to start the MCMC chain close to regions of high posterior probability since the parameters used to generate the sample of the predictive prior are known (the data were simulated; see the algorithm in section~\ref{alg}). Since the chain can be started immediately at the true values of the parameters, the burn in time is all but eliminated; the only exceptions being rare cases where the data is very unusual for the parameters which generated it.

\section{Selecting $\pi_{\mathcal{I}}$ and $\pi_{\mathcal{D}}$}\label{prior}

Having developed a tool to compare designs, we return to the problem at hand: Improving the design of OGTT tests.

We have discussed, in general, inference on OGTT data but we have as yet to fix the joint prior distribution that is to be used for $\theta_0, \theta_1$, $\theta_2$, and $G(0)$

We do not want misdiagnosed patients and we must make the best of available data to provide our inferences. An added difficulty is that the sample sizes involved are quite small. Testing repeated blood samples from a patient requires a significant amount of work from the laboratory staff, and requiring them to test a large number of blood samples is not reasonable (this may sometimes cause increased discomfort to the patient as well, although this is rare since the most common practice is to use a cannula). In practice the typical sample size is 3, although in some special research cases it may go up to 9. Consequently, priors must be chosen with a small sample size in mind. In particular, when performing inference, an informative prior is likely to overwhelm the data, and may lead to a diagnosis that is based mostly on the prior, rather than on the sample.

Assigning a prior distribution for inference which will serve for any patient is difficult. To avoid misdiagnosis, we must resort to a relatively vague prior. With this in mind, the priors chosen for $\pi_\mathcal{I}$ are the following:

$$
\theta_0\sim Gamma(2,1)
$$
$$
\theta_1\sim Gamma(2,1)
$$
$$
\theta_2\sim Gamma(10,1/20)\; 1{\hskip -2.5 pt}\hbox{I} \{\theta_2 > 0.16\}
$$
$$
G(0)\sim \mathcal{N}(80,100)\; 1{\hskip -2.5 pt}\hbox{I} \{G(0)\in [30,400] \}.
$$
We consider these priors to be vague since their regions of high probability extend well beyond any estimations performed with real patients.  $\theta_2$ has been truncated for mathematical reasons \citep[if $\theta_2$ is too small, then from the system of ODEs in section~\ref{model}, in (\ref{eq:dyn4}) and (\ref{eq:dyn5}), it will be possible for the glucose in the digestive system to begin with negative derivative, which is nonsense; see][]{campis} and $G(0)$ was truncated based on practical considerations: Any patient with an initial measurement anywhere near or below 30 or above 400$d/mL$
will not be tested but instead will be placed into emergency care (a preliminary, instant, fingerstick blood test is conducted,
for removal and immediate treatment of such cases). 

$\pi_I$ may be seen as an inadequate representation of our actual prior uncertainty, but using anything more informative can result in misdiagnosis of patients with unexpected glucose curves.  Since this prior is needed to analyze data arising from all patients, we must then settle for this relatively vague prior 

Now, if we set $\pi_{\mathcal{D}}$ equal to this vague $\pi_I$ our predictive prior will assign significant probability to regions that are not actually very likely scenarios. Our chosen design will therefore be tuned to take into consideration common situations as well as situations that occur infrequently, if they do. Our inference prior ($\pi_I$) was chosen for pragmatic reasons rather than based on an actual reflection of our uncertainty. For similar pragmatic reasons, it is not reasonable to use the same prior for design.

Moreover, as opposed to an informative inference prior, we do not expect the experimental design to have such a severe impact on misdiagnosis (this should be tested of course, but we know that OGTTs have been used successfully with a poor design for years), so we consider it less dangerous to use a more informative design prior.

Choosing a prior for design is also a difficult issue. One practical alternative is simply to pick a prior which represents the available data reasonably, use it to select a good design, and then compare it to arbitrary designs. There are several ways to pick the data in order to make comparisons, but one fair choice is to generate the data from $\pi_\mathcal{I}$. If the design appears to work well for data that is generated from the inference prior as well then we can conclude that this design is a good choice regardless of the prior that was used to generate it.

We propose using an \textit{extremely} informative prior for design. We have taken a sample of patients which represent typical scenarios, and have set our design prior to represent those specific patients. Our prior distribution gives an equal probability to each of the parameter combinations of these exact patients, and zero probability to anything else. This prior, of course, is not an adequate representation of our prior uncertainty either. If the design that is obtained when using our highly informative design prior proves to be useful for other patients as well, then this extreme prior will have served its purpose. In section \ref{resul} we will carefully examine how robust our results are, and whether our design proves suitable for other patients.

The reader might be inclined to take this approach of using a different prior for design and inference purposes as perhaps uncouth or strange. However, similar approaches have been studied in the context of reference priors, where the priors used for inference are different from the priors used for model selection, even when the context is the same \citep{pericchi}. Discussions of the use of different priors for design and inference - in different contexts - can also be found in \cite{scot} and also in \cite{stone}. Of note, the circumstances which lead us to the selection of different priors for inference and design are not actually very unusual; inference priors are often selected with high entropy in order to avoid overwhelming the information contained in the data, specially when dealing with small sample sizes. In such a case, the use of a different prior for design may be something to consider. The extreme case is seen when using improper (reference) priors for inference, wherein there is really no choice at all since design priors must be proper for $U(d)$ to be well-defined. In such a case, $\pi_\mathcal{I}$ and $\pi_\mathcal{D}$ \textit{must} be different.

\section{Implementation and Results}\label{resul}

The algorithm in section~\ref{alg} was used to select a design for OGTT diagnosis. In order to calculate $G(t)$, the forward map was solved numerically using the LSODA package for ordinary differential equations and an MCMC was used to sample from the posterior distribution using the t-walk package \citep{christen}.  The t-walk is an MCMC algorithm which is designed to adjust to continuous posterior distributions without tuning, which is particularly useful for our purposes since it means the MCMC does not have to be tuned separately for each patient. 

The utility function used was the negative mean squared error of $G(t)$, integrated over the curve from $t=0:00$ to $t=3:00$ hours. The choice was made so as not to attach significant preference to any particular time or parameter. The utility function is estimated by numerically estimating the integrated squared error for each element of a posterior sample and averaging across the estimators. One problem with this estimator is that it is only asymptotically unbiased rather than unbiased for finite sample size (see section~\ref{propt1yt2}) so a large $T_2$ was used. 

The selection of a design was done somewhat crudely, only comparing designs chosen with times at 15 minute intervals over a 2 hour period. A finer tuned selection would be significantly more expensive computationally, and it is not clear that it would be of much practical use since health professionals might not be able to take measurements at times which are specified with great precision while also keeping up with their other duties. 

In order to decide how many measurement times are required, comparisons are not done sequentially but allowed to be inconclusive if the decision cannot be made with a large $T_1$. "Large" in this case means 600. For such situations, where an experiment is to be performed several times, this number is interpretable; each element represents a simulated patient. If a decision cannot be reached with $T_1=n$ then this means that no difference is detectable when performing an OGTT test over a sample of $n$ patients. The number of measurements was deemed sufficient when adding another measurement resulted only in inconclusive comparisons.

The Python 2.7 programming language was used, running the t-walk MCMC algorithm for 1500 iterations for each patient. One such run takes between 5 and 10 seconds on an Intel processor running at 1.7GHz. To compare designs, a sample of 600 patients is taken for each, (unless one of these designs already has samples available from a previous comparison). One such comparison takes about 15 minutes. The full process is computationally intensive, but not unreasonably so, and can be parallelized for additional efficiency if needed.  In this particular case, considering 15 min intervals only, the full process took roughly 6 hours.

The resulting selection of times is $t=0:00, t=0:45, t=1:15, t=1:45$ and $t=2:00$ hours. In table~\ref{tab:1} we see the times from our newly proposed design next to the times from the conventional design which measures every hour. We also see a "Full" design that is sometimes used for research purposes. It is not practical to use this design in general, but it is used for validation purposes in section 8.1.2

\begin{table}
\begin{tabular}{|c|c|c|c|c|c|c|c|c|c|}
\hline 
mins & 0:00 & 0:15 & 0:30 & 0:45 & 1:00 & 1:15 & 1:30 & 1:45 & 2:00 \\
\hline 
Conventional & x & & & & x & & & & x \\ 
\hline
Proposed & x & & & x & & x & & x & x \\
\hline
Full & x & x & x & x & x & x & x & x & x\\
\hline
\end{tabular}
\\
\caption{Conventional times for glucose measurement in OGTTs and our proposed times. The "Full" times are used for validation purposes in section~\ref{conv}}
\label{tab:1}
\end{table}

\subsection{Validation}

\subsubsection{Comparison with arbitrary designs}

In order to check how robust this design is across varying data structures, the following experiment was performed: We selected sample sizes of 4, 5 and 6 data points (including measurement upon arrival). 100 designs were generated uniformly at random for each size. For each design a random ``patient '' was simulated, drawing $\vartheta$ from $\pi_{\mathcal{I}}$. For each simulated patient a sample was simulated for the random design and also for the proposed design. Inference was performed on each sample and the utility as described in section~\ref{resul} was estimated.

\begin{figure}
\begin{tabular}{c c c}
\includegraphics[height=3cm, width=4cm]{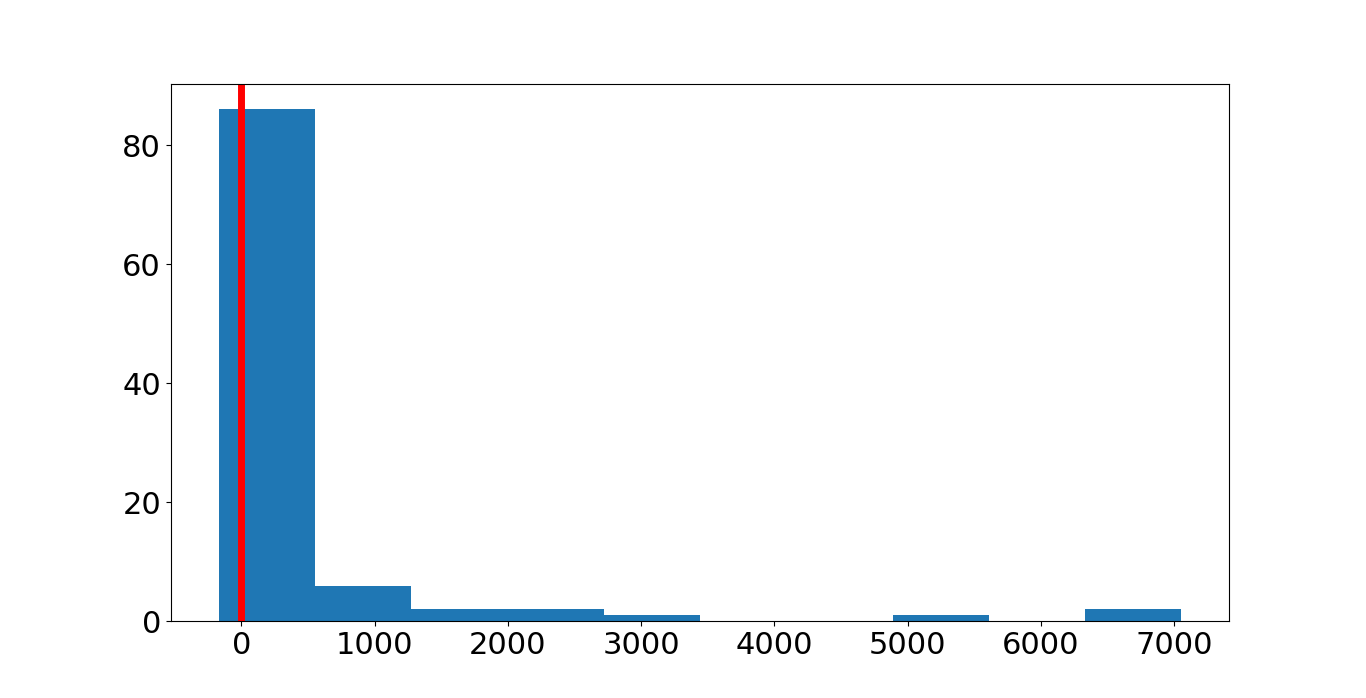} &
\includegraphics[height=3cm, width=4cm]{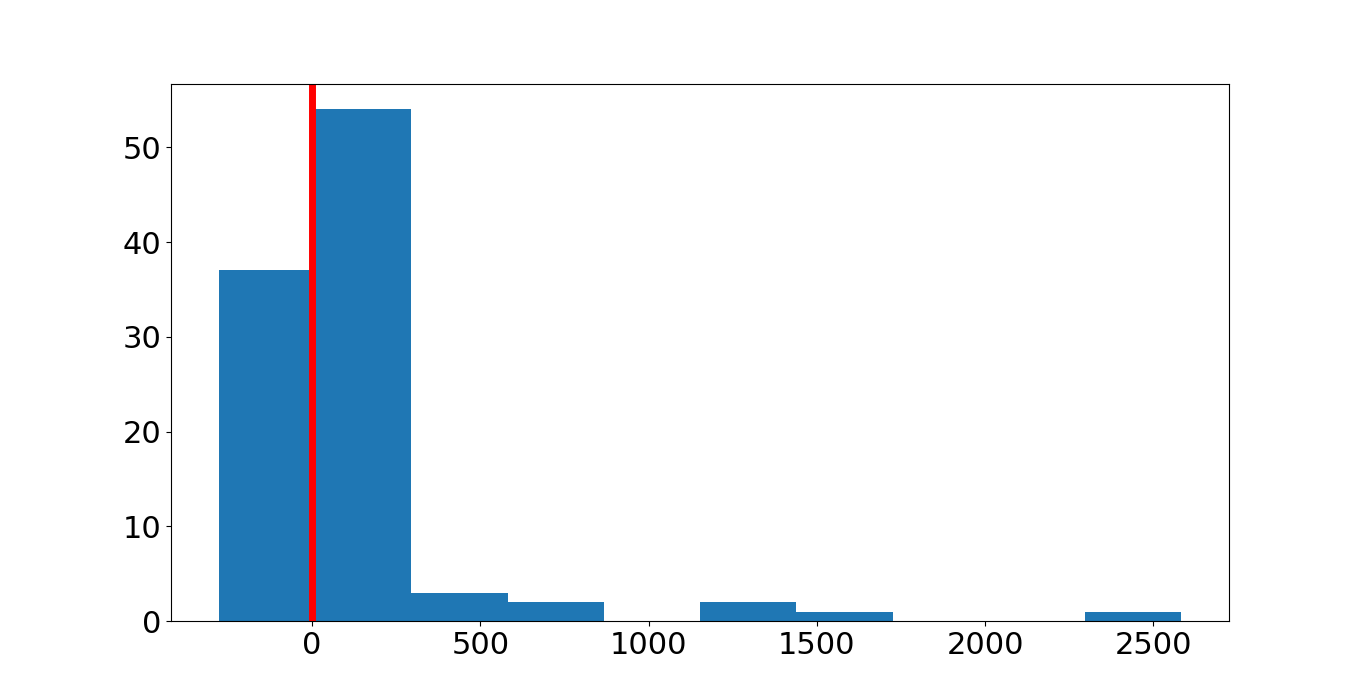} &
\includegraphics[height=3cm, width=4cm]{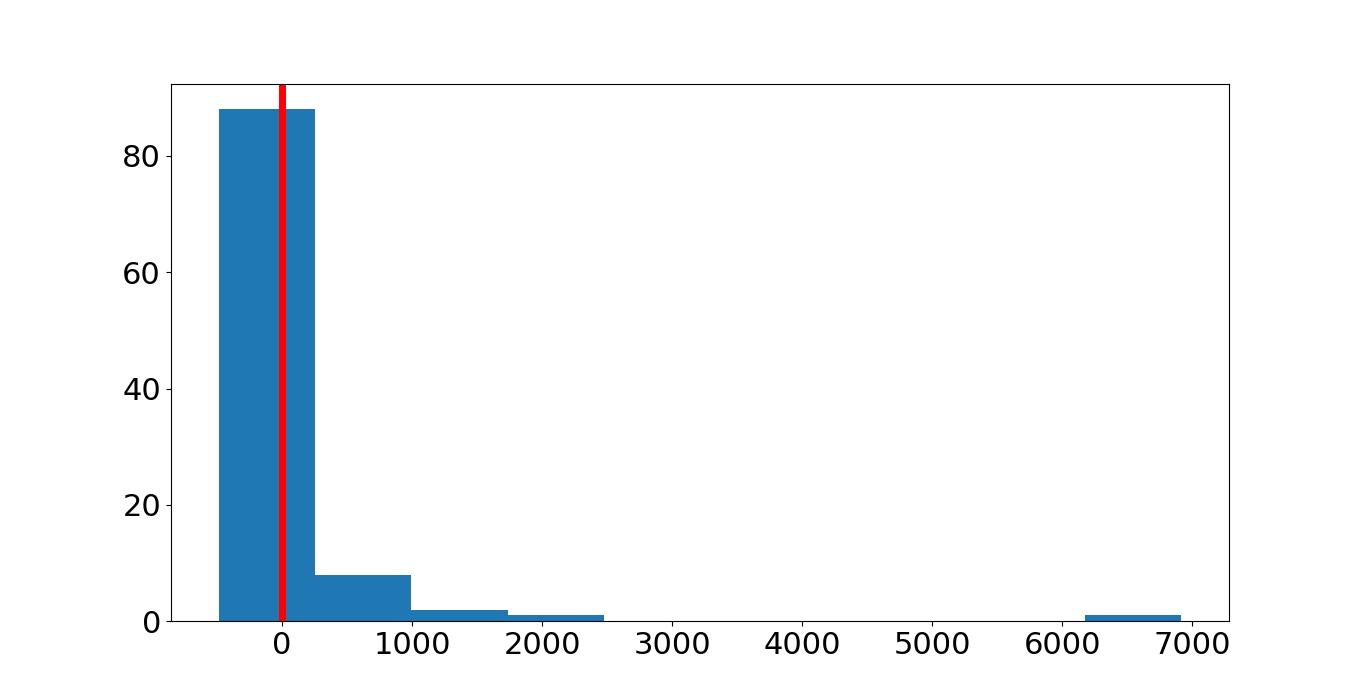} \\
(a) & (b) & (c)
\end{tabular}
\caption{Histograms of differences between the quality of our proposed design and of an arbitrary design on random data (arbitrary units). The vertical line indicates a difference of zero. The arbitrary design has one point less (a), the same number of points (b) and one more point (c) than our proposed design with 5 measuring points, seen in Table ~\ref{tab:1}. Note that, with the considered sample sizes, including bigger designs all of these histograms have right tails and none of them have a left tail. This means that our proposed design is never significantly worse than the arbitrarily chosen alternative, and is sometimes much better.}
\label{fig:differences}
\end{figure}

Figure~\ref{fig:differences} shows the histograms of the differences in utility between the arbitrary design and our suggested design for each sample size (utility of suggested design minus utlity of arbitrary design). We see a general trend: For most values of the parameters, the design does not make a very big difference in the quality of inference, thus the differences cluster around zero. All of the histograms have a right tail, and none of them have a left tail: For some values the design is more important; in these situations our design significantly outperforms the arbitrary design, even when the arbitrary design has a larger sample size. We can therefore conclude that our design does appear to be a generally good choice.

\subsubsection{Comparison with the conventional design} \label{conv}

While it is a very good sign that our design outperforms random designs, it is also important to compare our results with the conventional OGTT testing design which is actually used in practice. In the conventional design measurements are taken at $t={0,1,2}$ hours. This design has two fewer measurements than our proposed design so it is reasonable to expect that our design will be better for that reason alone, but it also means that the design is more costly. Quantifying the improvement over the classical design is therefore necessary to understand if and when this extra cost pays off.

To compare our new design to the conventional one, we have a sample of 17 real (healthy) patients, obtained by AM, for whom OGTT measurements were taken every 15 minutes, resulting in information that is significantly more complete than what is usually available from OGTT tests.  The conventional and proposed designs, as well as the full design were shown in table ~\ref{tab:1}.

In order to compare the two designs, the utility function must be estimated, but since these are real patients, the true value of the parameters is unknown. It is therefore not possible to estimate the expected utility with the precision which was used before, but a surrogate utility can be written which behaves similarly using the inference from the full data. The true utility function can be written as
$$
U(d) = - \int \int_0^3 (G_{\theta}(t)-G_{\hat{\theta}}(t))^2 dt ~ \pi_{\mathcal{I}}(\hat{\theta}|y,d) d\hat{\theta}.
$$
Since in this case the true parameters $\theta$ are unavailable we use their posterior distribution as calculated using the data from the full design. Our new surrogate utility is now
$$
\hat{U}(d)=- \int \int \int_0^3(G_{\theta}(t)-G_{\hat{\theta}}(t))^2 dt ~ \pi_{\mathcal{I}}(\hat{\theta}|y,d) d\hat{\theta} ~ \pi_{\mathcal{I}}(\theta |y_{f}) d\theta
$$
where $\pi_{\mathcal{I}}(\theta |y_{f})$ is the posterior distribution of the parameters $\theta$ using the full data $y_{f}$, that is, with measurements every 15min. This surrogate utility can be estimated using the available samples.

\begin{figure}
\includegraphics[height=9cm, width=13cm]{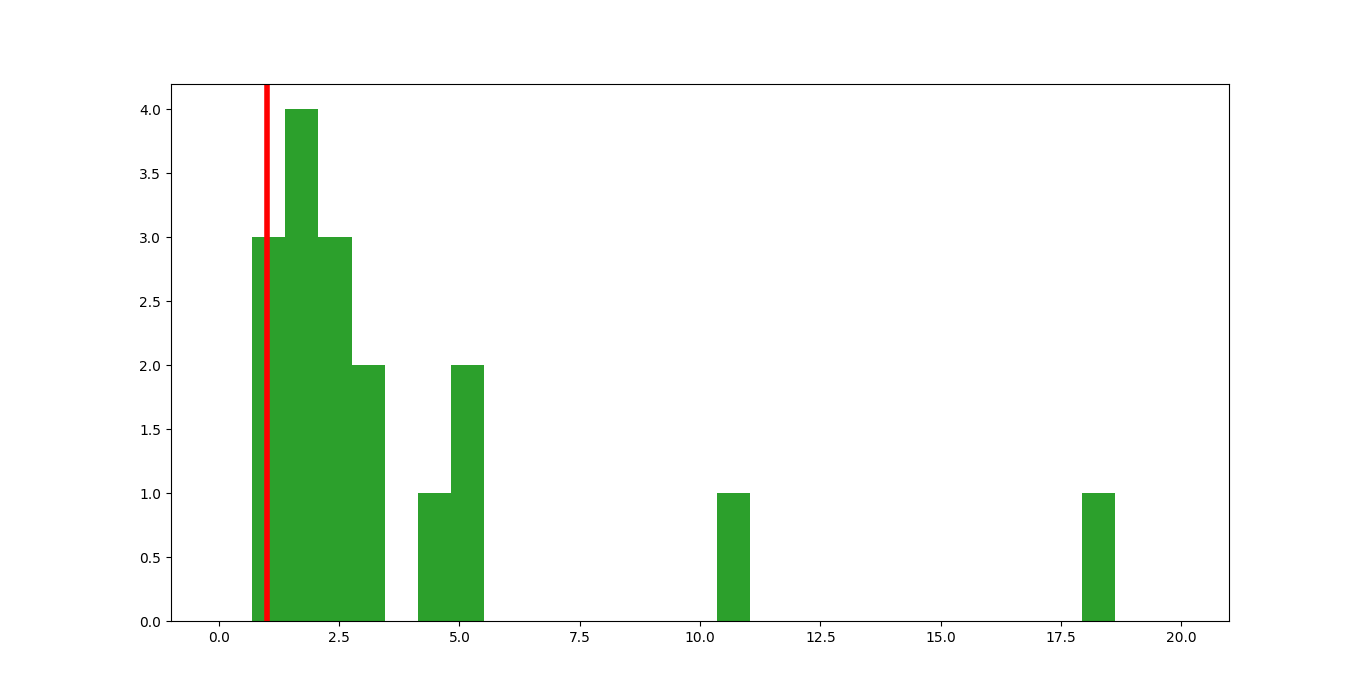}
\caption{Histogram of the quotients of the surrogate utility functions for 17 real patients using the conventional and proposed designs (conventional divided by proposed: All values are negative, so large quotients mean the conventional design yields larger errors). The vertical line indicates a quotient of 1.}
\label{fig:quotients}
\end{figure}

This was done for the available set of 17 patients and the estimates for the surrogate utilities were compared using the conventional design and using our proposal.  It is not surprising that the new design is better than the conventional design since, to start with, it has more measurements, but we want to know how much better.  In order to adequately represent the relative difference, a histogram of the \textit{quotients} of these utilities can be seen in figure~\ref{fig:quotients}. There are two patients for whom the utility of the conventional design outperforms the new one. For no patient did the new design result in an estimated utility of less than 82\% of the utility of the conventional design. For all other patients the new design outperforms the conventional design, usually by a factor of 2 or greater, and sometimes by a much wider margin.

As this example shows, the effect of choosing a better design can be dramatic. For the data tested, our proposed design has proven to be a significant improvement; raising the number of measurements from 3 to 5 achieved more than twice the utility for most patients.

\subsubsection{Simulation test to verify robustness}

\begin{figure}
    \includegraphics[height=9cm, width=13cm]{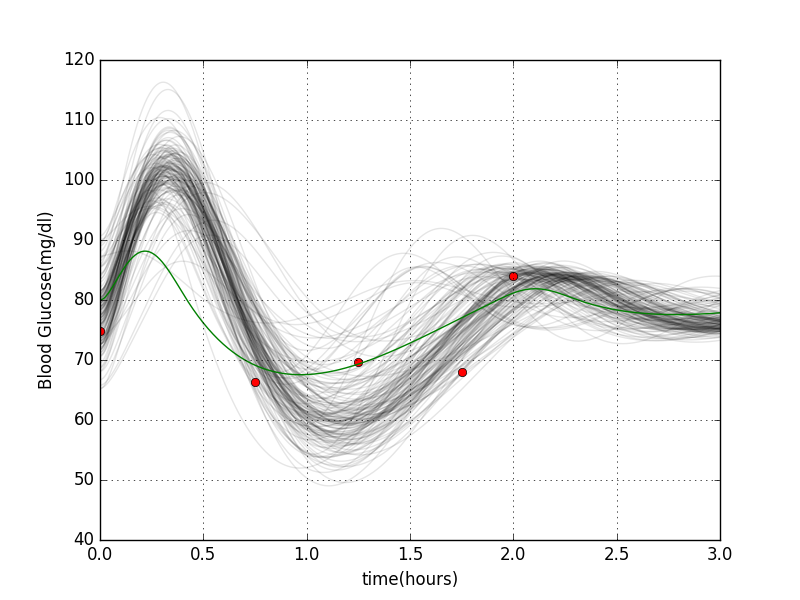}
    \caption{Simulated data for an extremely unusual situation where a patient's insulin response is 80 times stronger than the glucagon response. The true curve is in green and the simulated data in red. Although the data is extremely unusual, our design points yield the necessary information to obtain reasonable information about this strange behavior.}
    \label{fig:oddone}
\end{figure}

Another important test is to verify the behavior of this design when evaluating a particularly unusual set of data. Since the design was trained using typical scenarios, we should verify that highly unusual shapes can still be discovered. Data was simulated using a very strange set of parameters: $\theta_0=80$, $\theta_1=1$, $\theta_2=1.5$, $G_0=80$. This represents a patient whose insulin response is extremely violent, but whose glucagon production is not. The authors of this paper have never seen a patient like this one, with such a dramatic difference between the production of the two hormones. Data was simulated using the proposed design and a posterior sample can be seen in figure \ref{fig:oddone} with the true curve seen in green. As we can see, while the inference is imperfect, the design still works reasonably well in performing inference even on this extremely strange patient.

\section{Discussion}

In this paper the model proposed in Section~\ref{sec.dynmodel} was used to suggest a better analysis and improved sampling protocols for Oral Glucose Tolerance Tests. The main objective was to use the model to redesign the OGTT test in a way that improves the quality of the information gathered. The chosen technique to achieve these purposes was Bayesian experimental design, in order to find an alternative set of times at which to perform the glucose measurements on the patient.

Although some techniques for Bayesian experimental design already exist, the specific properties of this problem lead us towards developing a different tool for comparison of experimental designs, which is computationally intensive, but which provides a finer control of uncertainty in the design process itself.

We used this new tool to select a design for the OGTT and the resulting choice was compared favorably both to the classical design (with real data) and to hypothetical arbitrary designs.
The result is very promising and may lead to improved diagnosis techniques for patients who are at risk of type 2 diabetes.

From a mathematical perspective, there remains an issue regarding the algorithm for comparing designs, in those cases when a decision should be forced (by increasing $T_1$). In section~\ref{decide} we briefly discuss the notion of sequential comparisons in cases where the initial test proves inconclusive. While this was not necessary in our case, in most other cases the value of $T_1$ will not be easily interpretable. If our method is to be generally applicable, an efficient algorithm for sequential testing should be developed. Although we performed some numerical experiments in sequential design, we have not come to any clear conclusions regarding how to do it efficiently.

The most innovative and potentially controversial issue in this paper is of course the explicit use of two separate priors over the parameter space, one for design and one for inference purposes. In section~\ref{prior} we have discussed some pragmatic reasons why this may be a desirable - and in some cases necessary- option, but it may be possible to treat the subject more formally. Prior selection is -after all- a decision, and it may be possible to frame this instance of prior selection in the context of decision theory. The use of decision theoretic constructions to select priors has been studied in the context of reference analysis \citep[see for example][]{bernardo}
and a similar approach could shed light on this context as well.

\medskip
\bibliography{PaperDesign}
\end{document}